\documentclass[lettersize,journal]{IEEEtran}
\usepackage{amsmath,amsfonts}
\usepackage{algorithmic}
\usepackage{algorithm}
\usepackage{array}
\usepackage{subfig}
\usepackage[automake, acronym]{glossaries-extra}
\usepackage{textcomp}
\usepackage{accents}
\usepackage{stfloats}
\usepackage{xcolor}
\usepackage{url}
\usepackage{verbatim}
\usepackage{enumitem}
\usepackage{chronology}
\usepackage{graphicx}
\usepackage{cite}
\usepackage{multirow}
\usepackage{tikz}
\usepackage{makecell}
\usepackage{float}
\usetikzlibrary{arrows,positioning,decorations.pathreplacing}
\usetikzlibrary{fit}
\usepackage{xcolor}
\definecolor{blue}{rgb}{0,0,0}
\definecolor{red}{rgb}{0,0,0}
\makeglossaries 
\begin{document}

\title{Efficient Data-Driven Model Predictive Control for Demand Response of Commercial Buildings}

\author{Marie-Christine Paré, Vasken Dermardiros, and Antoine Lesage-Landry,~\IEEEmembership{Member,~IEEE} \thanks{This work was funded by IVADO, Mitacs, and the National Science and Engineering Research Council of Canada (NSERC).}
\thanks{M-C. Paré and A. Lesage-Landry are with the Department of Electrical Engineering, Polytechnique Montreal, MILA \& GERAD, Montréal, QC, Canada, H3T 1J4. e-mail: \texttt{marie-christine.pare@polymtl.ca}, \texttt{antoine.lesage-landry@polymtl.ca}}
\thanks{V. Dermardiros is with BrainBox AI, Montréal, QC, Canada, H3A 2L1. e-mail: \texttt{vdermardiros@gmail.com}}}
\markboth{IEEE TRANSACTIONS ON CONTROL SYSTEMS TECHNOLOGY,~Vol.~X, No.~X, December~2023}{ \MakeLowercase{Paré, Dermardiros, and Lesage-Landry}: Efficient Data-Driven MPC for Demand Response of Commercial Buildings}
\maketitle


\newglossaryentry{rtu}{
  type=\acronymtype,
  name={RTU},
  description={RTU},
  first={rooftop unit (RTU)},
  plural={RTUs},
  firstplural={rooftop units (RTUs)}
}

\newglossaryentry{icrnn}{
  type=\acronymtype,
  name={ICRNN},
  description={input convex recurrent neural network},
  first={input convex recurrent neural network (ICRNN)},
  plural={ICRNNs},
  firstplural={input convex recurrent neural networks (ICRNNs)}
}

\newglossaryentry{mpc}{
  type=\acronymtype,
  name={MPC},
  description={model predictive control},
  first={model predictive control (MPC)},
  plural={MPCs},
  firstplural={model predictive controls (MPCs)}
}

\newglossaryentry{lstm}{
  type=\acronymtype,
  name={LSTM},
  description={LSTM},
  first={long short-term memory (LSTM)},
  plural={LSTMs},
  firstplural={long short-term memories (LSTMs)}
}

\newglossaryentry{rnn}{
  type=\acronymtype,
  name={RNN},
  description={recurrent neural network},
  first={recurrent neural network (RNN)},
  plural={RNNs},
  firstplural={recurrent neural networks (RNNs)}
}

\newglossaryentry{nn}{
  type=\acronymtype,
  name={NN},
  description={neural network},
  first={neural network (NN)},
  plural={NNs},
  firstplural={neural networks (NNs)}
}

\newglossaryentry{hvac}{
  type=\acronymtype,
  name={HVAC},
  description={heating, ventilation, and air conditioning},
  first={heating, ventilation, and air conditioning (HVAC)},
  plural={heating, ventilation, and air conditioning systems},
  firstplural={heating, ventilation, and air conditioning systems (HVAC)}
}

\newglossaryentry{icnn}{
  type=\acronymtype,
  name={ICNN},
  description={input convex neural network},
  first={input convex neural network (ICNN)},
  plural={ICNNs},
  firstplural={input convex neural networks (ICNNs)}
}

\newglossaryentry{ficnn}{
  type=\acronymtype,
  name={FICNN},
  description={Fully Input Convex Neural Network},
  first={Fully Input Convex Neural Network (FICNN)},
  plural={Fully Input Convex Neural Networks},
  firstplural={Fully Input Convex Neural Networks (FICNNs)}
}

\newglossaryentry{picnn}{
  type=\acronymtype,
  name={PICNN},
  description={Partially Input Convex Neural Network},
  first={Partially Input Convex Neural Network (PICNN)},
  plural={Partially Input Convex Neural Networks},
  firstplural={Partially Input Convex Neural Networks (PICNNs)}
}

\newglossaryentry{nem}{
  type=\acronymtype,
  name={NEM},
  description={Australian National Electricity Market},
  first={Australian National Electricity Market (NEM)},
  plural={Australian National Electricity Markets},
  firstplural={Australian National Electricity Markets (NEMs)}
}

\newglossaryentry{fcas}{
  type=\acronymtype,
  name={FCAS},
  description={Frequency Control Ancillary Services},
  first={Frequency Control Ancillary Services (FCAS)},
  plural={Frequency Control Ancillary Services},
  firstplural={Frequency Control Ancillary Services (FCAS)}
}

\newglossaryentry{dr}{
  type=\acronymtype,
  name={DR},
  description={demand response},
  first={demand response (DR)},
  plural={DRs},
  firstplural={remand responses (DRs)}
}

\newglossaryentry{iso}{
  type=\acronymtype,
  name={ISO},
  description={independent system operator},
  first={independent system operator (ISO)},
  plural={Independent System Operators},
  firstplural={Independent System Operators (ISOs)}
}

\newglossaryentry{armax}{
  type=\acronymtype,
  name={ARMAX},
  description={Autoregressive-Moving-Average with Exogenous Inputs},
  first={Autoregressive-Moving-Average with Exogenous Inputs (ARMAX)},
  plural={Autoregressive-Moving-Averages with Exogenous Inputs},
  firstplural={Autoregressive-Moving-Averages with Exogenous Inputs (ARMAXs)}
}

\newglossaryentry{vns}{
  type=\acronymtype,
  name={VNS},
  description={Variable Neighborhood Search},
  first={Variable Neighborhood Search (VNS)},
  plural={Variable Neighborhood Searches},
  firstplural={Variable Neighborhood Searches (VNSs)}
}

\newglossaryentry{mads}{
  type=\acronymtype,
  name={MADS},
  description={Mesh Adaptive Direct Search},
  first={Mesh Adaptive Direct Search (MADS)},
  plural={Mesh Adaptive Direct Searches},
  firstplural={Mesh Adaptive Direct Searches (MADSs)}
}

\newglossaryentry{dfo}{
  type=\acronymtype,
  name={DFO},
  description={derivative-free optimization},
  first={derivative-free optimization (DFO)},
  plural={derivative-free optimizations},
  firstplural={derivative-free optimizations (DFOs)}
}

\newglossaryentry{oat}{
  type=\acronymtype,
  name={OAT},
  description={outdoor air temperature},
  first={outdoor air temperature (OAT)},
  plural={outdoor air temperatures},
  firstplural={outdoor air temperatures (OATs)}
}

\newglossaryentry{iat}{
  type=\acronymtype,
  name={IAT},
  description={indoor air temperature},
  first={indoor air temperature (IAT)},
  plural={indoor air temperatures},
  firstplural={indoor air temperatures (IATs)}
}

\newglossaryentry{ghi}{
  type=\acronymtype,
  name={GHI},
  description={GHI},
  first={global horizontal irradiance (GHI)},
  plural={GHIs},
  firstplural={global horizontal irradiances (GHIs)}
}

\newglossaryentry{mip}{
  type=\acronymtype,
  name={MIP},
  description={Mixed-Integer Program},
  first={Mixed-Integer Program (MIP)},
  plural={Mixed-Integer Programs},
  firstplural={Mixed-Integer Programs (MIPs)}
}

\newglossaryentry{milp}{
  type=\acronymtype,
  name={MILP},
  description={Mixed-Integer Linear Program},
  first={Mixed-Integer Linear Program (MILP)},
  plural={Mixed-Integer Linear Programs},
  firstplural={Mixed-Integer Linear Programs (MILPs)}
}

\newglossaryentry{cpr}{
  type=\acronymtype,
  name={CPR},
  description={critical peak rebate},
  first={critical peak rebate (CPR)},
  plural={critical peak rebates},
  firstplural={critical peak rebates (CPRs)}
}

\newglossaryentry{tou}{
  type=\acronymtype,
  name={TOU},
  description={Time-of-use},
  first={time-of-use (TOU)}}

\newglossaryentry{cb}{
  type=\acronymtype,
  name={CB},
  description={commercial building},
  first={commercial building (CB)},
  plural={CBs},
  firstplural={commercial buildings (CBs)}
}

\newglossaryentry{da}{
  type=\acronymtype,
  name={DA},
  description={day-ahead},
  first={day-ahead (DA)},
}

\newglossaryentry{sp}{
  type=\acronymtype,
  name={SP},
  description={day-ahead},
  first={set-point (SP)},
}

 \newglossaryentry{mimo}{
  type=\acronymtype,
  name={MIMO},
  description={mimo},
  first={multiple-input and multiple-output (MIMO)},
}

\newglossaryentry{arx}{
  type=\acronymtype,
  name={ARX},
  description={autoregressive exogenous input (ARX)},
  first={autoregressive exogenous input (ARX)},
}

\newglossaryentry{mcp}{
  type=\acronymtype,
  name={MCP},
  description={market clearing price (MCP)},
  first={market clearing price (MCP)},
    plural={MCPs},
  firstplural={market clearing prices (MCPs)}
}

\newglossaryentry{rmse}{
  type=\acronymtype,
  name={RMSE},
  description={root mean square error (RMSE)},
  first={root mean square error (RMSE)},
    plural={RMSEs},
  firstplural={root mean square errors (RMSEs)}
}

\newglossaryentry{mae}{
  type=\acronymtype,
  name={MAE},
  description={mean absolute error (MAE)},
  first={mean absolute error (MAE)},
    plural={MAEs},
  firstplural={mean absolute error (MAEs)}
}

\newglossaryentry{admm}{
  type=\acronymtype,
  name={ADMM},
  description={alternating direction method of multipliers (ADMM)},
  first={alternating direction method of multipliers (ADMM)},
    plural={ADMM},
  firstplural={alternating direction method of multipliers}
}
\begin{abstract}
\Gls{mpc} has been shown to significantly improve the energy efficiency of buildings while maintaining thermal comfort. \textcolor{blue}{Data-driven approaches like neural networks can facilitate system modelling.} However, such approaches are generally nonconvex and result in computationally intractable optimization problems. In this work, we design a readily implementable energy management method for small commercial buildings. We then leverage our approach to formulate a real-time demand bidding strategy. We propose a data-driven and mixed-integer convex \gls{mpc} which is solved via derivative-free optimization given a limited computational time of 5 minutes to respect operational constraints. We consider rooftop unit heating, ventilation, and air conditioning systems with discrete controls to accurately model the operation of most commercial buildings. Our approach uses an input convex recurrent neural network to model the thermal dynamics. We apply our approach \textcolor{blue}{to} several \gls{dr} settings, including a demand bidding, a time-of-use, and a critical peak rebate program. Controller performance is evaluated on a state-of-the-art building simulation. The proposed approach improves thermal comfort while reducing energy consumption and cost through \gls{dr} participation, when compared to other data-driven approaches or a set-point controller.
\end{abstract}


\glsresetall
\section{Introduction}
\IEEEPARstart{T}{he} built environment stands as one of the most energy-intensive sectors, and as one of the main contributors to climate change. The buildings sector alone accounts for 30\% of global energy consumption and contributes to 28\% of total greenhouse gas emissions~\cite{gonzalez2022review}. 
In the U.S., \glspl{cb} alone consume 46\% of the energy used in the building sector. A substantial portion of this energy is allocated to operating their \gls{hvac} systems, which constitute 47\% of their total energy consumption~\cite{allouhi2015energy}.
Inefficient \gls{hvac} operation, due to the lack of adequate control, often results in significant energy waste~\cite{HVAC_modelling}. In the U.S., over 90\% of \glspl{cb} fall into the small- to medium-sized category and are predominantly equipped with packaged \glspl{rtu}. It is estimated that the majority of these buildings still rely on rudimentary controllers, such as ruled-based or \gls{sp}-based. 

\Gls{mpc} has demonstrated its ability to improve thermal comfort while achieving energy and cost savings through efficient \gls{hvac} operation~\cite{HVAC_modelling}. Despite extensive research and practical case studies, \textcolor{blue}{adoption of the technology within the \glspl{cb} industry remains limited, primarily due to challenges in automating design, tuning, and implementation}. The quality of an \gls{mpc} solution depends on the model's accuracy and the ability to solve the optimization problem efficiently and globally at its core. \textcolor{blue}{Physics-based approaches produce reliable, interpretable models but require significant development effort~\cite{review_MPC}}. In contrast, data-driven approaches promote wide-scale implementation, because they do not require understandings of the system dynamics. This translates into a more transferable approach and lower development costs~\cite{HVAC_modelling}. Modelling the thermal dynamics of a building can be a challenging task due to their highly nonlinear nature~\cite{MORK2022}. To address these challenges, many have adopted an approach based on \textcolor{blue}{\glspl{rnn}} because of their high modelling accuracy \textcolor{blue}{for dynamic systems}~\cite{wang2019data}. Although \textcolor{blue}{\gls{rnn}}-based \gls{mpc} have been applied to \glspl{cb} in~\cite{BBAI}, this results in a nonconvex optimization problem which is computationally intractable. \textcolor{blue}{In practice, modelling accuracy is often sacrificed for control tractability, favoring linear models despite their limitations~\cite{wang2019data}.}

As part of the largest consumers, \glspl{cb} can play a significant role in \gls{dr}. Their thermal mass provides inherent flexibility, enabling their participation in \gls{dr} events without compromising thermal comfort. Achieving accurate real-time control of \glspl{cb} \textcolor{blue}{can} unlock a large amount of flexibility, thereby enhancing power system \textcolor{blue}{resiliency}, efficiency, and \textcolor{blue}{reliability} under high renewable energy penetration, while generating savings for participants~\cite{Flexibility_categorization}.

In this work, we design a readily implementable energy management method for demand response of small \glspl{cb} equipped with \gls{rtu}-\gls{hvac}. Our approach is based on a convex \gls{mpc} and uses an \gls{icrnn} to model the thermal dynamics in a data-driven fashion. We consider discrete controls to accurately model the operation of most \gls{rtu}-\gls{hvac} \glspl{cb}. We propose a resolution method based on a \gls{dfo} solver. \textcolor{blue}{The convexity of our approach promotes more efficient computation compared to a conventional RNN-based MPC, as the solver avoids local minima.} Then, we formulate a \gls{dr} strategy where \glspl{cb} can bid flexibility in a real-time market and present numerical case studies in which various \gls{dr} programs are discussed, showcasing the full potential of our method. We test our method on a Modelica~\cite{WetterZuoNouiduiPang2014} simulation of a 2-zone \gls{cb}, a state-of-the-art physics-based simulation.

We now review the related literature on input convex \glspl{nn} and \gls{mpc}-based bidding strategy for \glspl{cb}. Efforts have been made to develop a class of input convex \glspl{nn} that can be easily embedded in an optimization framework like \gls{mpc}. First introduced by~\cite{Amos}, \glspl{icnn} have constraints on their architecture that guarantee the \gls{nn}'s convexity in its inputs, thus permitting efficient global optimization. In~\cite{chen2019}, the class of convex \glspl{nn} is extended to include the \gls{icrnn}. The authors of~\cite{chen2019} also formally \textcolor{blue}{prove} the representation ability and efficiency of both \glspl{icnn} and \glspl{icrnn}. In their case studies, \textcolor{blue}{a real-time \gls{hvac} control experiment is presented on a simulation of a large-scale office building}. The \gls{icrnn}-based \gls{mpc} \textcolor{blue}{achieves} the largest energy reduction when compared to the ones based on a nonconvex \gls{rnn} or a linear model. In~\cite{bunning_MPC}, a real-life case study considering an apartment room is presented and \textcolor{blue}{shows} that an \gls{icnn}-based MPC successfully keeps the temperature within the comfort zone most of the time. In~\cite{BUNNING_2022}, the authors \textcolor{blue}{of~\cite{bunning_MPC} extend} their previous work to compare various convex and data-driven \gls{mpc} approaches, namely a thermal model based on a random forest, a physics-informed ARMAX, and an \gls{icnn}. Contrarily to previous work~\cite{chen2019,bunning_MPC, BUNNING_2022} in this literature stream, we here consider \textcolor{blue}{discrete \gls{hvac} controls and solve the problem as a mixed-integer optimization problem}. We then utilize our approach for several \gls{dr} applications. Discrete control models promote minimal building upgrades and hence increase its potential for the practical implementation of our method.

\gls{mpc}-based methods have been leveraged for demand bidding of \glspl{cb}. In~\cite{Exp_demo}, the authors \textcolor{blue}{present} an experimental demonstration where \glspl{cb}' \gls{hvac} systems are used to provide ancillary frequency regulation services in the Swiss energy market. Participants acquire energy in the \gls{da} market, which serves as a baseline for the following day, and submit flexibility bids accordingly. The baseline and bid quantities are determined by solving a robust optimization problem, and a standard \gls{mpc} is used to operate \gls{hvac} systems and revise bids in the intraday market. In~\cite{bid_MPC1}, a method for real-time coordination of building consumption and energy scheduling of power systems is proposed. This method utilizes an \gls{mpc} to plan the operation of \gls{hvac} systems and battery energy storage, as well as to generate demand bid curves. In~\cite{NYC_MPC}, an \gls{mpc} framework for the \gls{da} market enables \glspl{cb} to submit bids for demand reductions. The control architecture consists of two \textcolor{blue}{sequential} levels: the first level determines the building's daily participation in the market, and the second level handles real-time operations. These works lack a precise model of the thermal dynamics, as they depend solely on a linear model and are generally tailored to a specific market. We address these limitations by formulating a new adaptable \gls{mpc}-based bidding strategy utilizing a discrete \gls{icrnn} \gls{cb} model. 

Our specific contributions are as follows.
\begin{itemize}
\item [\tiny$\bullet$] We formulate a mixed-integer piecewise-linear and data-driven \gls{mpc} \textcolor{blue}{for cooling operations} of \gls{rtu}-based small \glspl{cb}  (Section~\ref{base_controller}). 
\item [\tiny$\bullet$] We leverage a precise discrete model to formulate an adaptable real-time \gls{dr} bidding strategy (\textcolor{red}{Section~\ref{demand_bidding_modelling}}).
\item [\tiny$\bullet$] We present numerical case studies showcasing several \gls{dr} settings (Sections \ref{demand_bidding}$-$\ref{Critical_peak_rebate}). 
\end{itemize}

We detail our modelling approach and formulate \gls{rtu}-\gls{hvac} control problems in Section~\ref{modelling_and_problem}. In Section~\ref{case_study}, we present our case studies wherein we assess the performance of our proposed approach, covering inference in Section~\ref{model_training} and control in Section~\ref{Controller_performances}. We conclude in Section~\ref{conclusion}. 

\section{Modelling and Problem Formulation}\label{modelling_and_problem}
We now present the considered building setting, then introduce the thermal dynamics model. We conclude with \glspl{mpc} for both a base controller and a demand bidding strategy.
\subsection{System Description}\label{system_description}
We focus on a small \gls{cb} served by \glspl{rtu}, divided into $n_{\text{z}} \in \mathbb{N}$ zones, where $n_{\text{z}}$ usually ranges from 1 to 3\textcolor{blue}{~\cite{smb_2012}}. Each zone is equipped with its own thermostat and its associated \gls{rtu} receiving controls every 5 minutes. A typical \gls{rtu} comprises $n_{\text{u}}  \in \mathbb{N}$ components, such as electric heating and cooling coils with two intensities each, as well as a fixed-speed fan. Each of these components is controlled using \textsc{on}/\textsc{off} actuation. \textcolor{blue}{The reduced control space for cooling operations can be summarized as two intensities of cooling, $u_{\text{c}1}\in \{0,1\}$ for intensity 1 and $u_{\text{c}2}\in \{0,1\}$ for intensity 2, and fan ventilation $u_{\text{f}}\in \{0,1\}$. Let $\mathbf{u}_{\text{RTU}}  = \left[u_{\text{c}2}, u_{\text{c}1}, u_{\text{f}} \right]^{\top}$ be the control variables vector. The set $\mathcal{U}= \{[0,0,0,0,0 ]^{\top}, [0,0,1,0,0]^{\top}, [0,1,1,0,0]^{\top}, $ $ [1,1,1,0,0]^{\top}\}$} is the set of valid \textcolor{blue}{reduced} controls for one \gls{rtu}, where $\mathbf{u}_{\text{RTU}} \in \mathcal{U}$. For the remainder of this work, we consider \textcolor{blue}{the possibility of having more than one \gls{rtu}}, such that the number of controls is $n_{\text{z}}n_{\text{u}}$. Let $\mathbf{u}\in \mathcal{U}^{n_{\text{z}}}$ represent the controls for all $n_{\text{z}}$ \glspl{rtu} concatenated together. Let $\mathbf{p}^{\text{c}} \in \mathbb{R}^{n_{\text{z}}n_{\text{u}}}$ denote the component power rating of the \glspl{rtu} collected within a vector.


\subsection{Building Thermal Modelling}\label{thermal_modelling}  
\begin{figure}[b]
    \centering
    \includegraphics[clip=true, trim= {0.5cm 21.85cm 10cm 0cm}, scale=0.60]{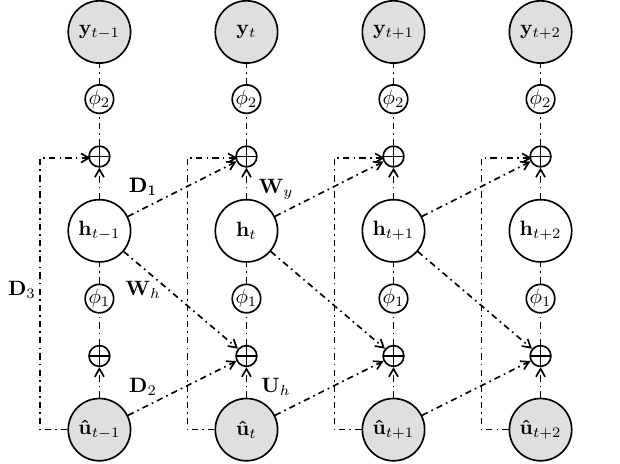}
    \caption{Architecture of an \gls{icrnn}.}
    \label{fig:icrnn}
\end{figure}

We base our approach on the work of~\cite{chen2019} to fit a convex and data-driven model of the building's thermal dynamics with an \gls{icrnn}. The \gls{icrnn} architecture is presented in Figure~\ref{fig:icrnn}. \Glspl{rnn} are a class of \glspl{nn} that carry a state to represent the information seen in previous time steps with a memory window length $w\in\mathbb{N}$, enabling them to perform accurate multi-step predictions~\cite{FAN_2019, chen2019}. Let $\hat{\mathbf{u}}_{t}=\left[ \mathbf{u}_{t}, -\mathbf{u}_{t}\right]$ represent the extended inputs, and let $\mathcal{\hat{U}}$ denote the set of extended valid controls. By constraining the architecture of an \gls{rnn} during training, we can guarantee the convexity of the resulting scalar function with respect to its inputs $\hat{\mathbf{u}}_{t}$, $\forall t$. As a result, the trained \gls{icrnn} can be embedded \textcolor{blue}{into} an optimization problem which can then be solved efficiently to global optimality. Such \gls{nn} is described by~(\ref{hidden_state})$-$(\ref{outputs}). In our context, it maps input $\hat{\mathbf{u}}_{t} \in \mathcal{\hat{U}} $ to output $\mathbf{y}_{t}\in \mathbb{R}^{n_{\text{z}}}$ with hidden state $\mathbf{h}_{t}: \textcolor{blue}{\mathcal{\hat{U}}} \times \mathbb{R}^{n_{\text{h}}} \mapsto \mathbb{R}^{n_{\text{h}}}$, where $n_{\text{h}}\in \mathbb{N}$ is the number of hidden units, with the parameters $\mathbf{U}_{\text{h}}$, \textcolor{blue}{$\mathbf{W}_{\text{h}}$}, $\mathbf{P}_{2}$, $\mathbf{b}_{\text{h}}$, $\mathbf{W}_{\text{y}}$, $\mathbf{P}_{1}$, $\mathbf{P}_{3}$, and $\mathbf{b}_{\text{y}}$\textcolor{blue}{:}
\begin{align}
    \mathbf{h}_t &= \phi_1\left(\mathbf{U}_{\text{h}} \mathbf{\hat{u}}_{t} + \mathbf{W}_{\text{h}} \mathbf{h}_{t-1} + \mathbf{P}_2 \mathbf{\hat{u}_{t-1}} + \mathbf{b}_{h}\right)  \label{hidden_state} \\ 
    \mathbf{y}_{t} &= \phi_2\left(\mathbf{W}_{\text{h}}\mathbf{h}_{t} + \mathbf{P}_{1} \mathbf{h}_{t-1} + \mathbf{P}_{3} \mathbf{\hat{u}_{t}} + \mathbf{b}_{\text{y}} \right). \label{outputs} 
\end{align}
The sufficient conditions for the convexity of the network defined by~(\ref{hidden_state})$-$(\ref{outputs}) are that its parameters  $\mathbf{U}_{\text{h}}$, \textcolor{blue}{$\mathbf{W}_{\text{h}}$}, $\mathbf{P}_{2}$, $\mathbf{W}_{\text{y}}$, $\mathbf{P}_{1}$, and $\mathbf{P}_{3}$ are non-negative and its activation functions $\phi_{1}$ and $\phi_{2}$ are convex and non-decreasing (e.g., ReLU or LeakyReLU). The resulting trained-\gls{nn} is convex and non-decreasing over $\hat{\mathbf{u}}_{t}$, but its \textcolor{blue}{output} can be decreasing or non-decreasing with respect to $\mathbf{u}_{t}$.


\subsection{Model Predictive Control}\label{model_predictive_control}
We now introduce our control problems. We begin by presenting our base controller, followed by our demand bidding strategy.
\subsubsection{Base Controller}\label{base_controller} The \gls{cb} energy management problem is formulated as a rolling horizon optimization problem. The objective is to minimize electrical operational costs, i.e., the cost of energy use and peak demand associated with the \glspl{rtu}, while maintaining thermal comfort and satisfying equipment requirements over the prediction horizon $T\in \mathbb{N}$, in all $n_{\text{z}}$ zones. For all \textcolor{blue}{$t\in\{1,2,\dots\}$}, let $\mathcal{T}_{t}=\{t, \dots t+T\}$ be the set of \gls{mpc} intervals. Let $\boldsymbol{\theta}^{t} \in \mathbb{R}^{n_{\text{z}}}$ be the \gls{iat}, and $\Delta \boldsymbol{\theta}^{t} = \boldsymbol{\theta}^{t}-\boldsymbol{\theta}^{t-1}$ the \gls{iat} variations between two sampling steps $\Delta t>$~0, in all $n_{\text{z}}$ zones. 
The \gls{mpc} problem is presented in~(\ref{obj})$-$(\ref{c_domain}) and described next.\textcolor{blue}{
\begin{alignat}{2}
\underset{\mathbf{u}^{k}, \text{ } k\in\mathcal{T}_{t}}{\text{min}} &  \sum_{k \in \mathcal{T}_{t}} \lambda_{\text{E}}^{k} E^{k} + \lambda_{\text{P}} \underset{k \in \mathcal{T}_{t}}{\max}\left\{ {\mathbf{u}^{k}}^\top  \mathbf{p}^{\text{c}} \right\} \label{obj}  \\  
\text{ s.t. } & \Delta \boldsymbol{\theta}^{k} = f_{\text{ICRNN}}\left(\mathbf{x}^{k-w}, \dots, \mathbf{x}^{k} \right), \forall k \in \mathcal{T}_{t} \label{dynamics} \\ 
 & \boldsymbol{\theta}^{k-1} + \Delta \boldsymbol{\theta}^{k} \leq \Bar{\boldsymbol{\theta}}^{k}, \forall k \in \mathcal{T}_{t}  \label{upper_temp_bound}\\ 
 &\sum_{\tau=k-\rho}^{k-1}\max\left\{\mathbf{u}^{\tau-1}-\mathbf{u}^{\tau}, \mathbf{0}\right\} \leq  1-\mathbf{u}^{k}, \forall k \in \mathcal{T}_{t} \label{lock-out} \\
 & E^{k} =  {\mathbf{u}^{k}}^\top  \mathbf{p}^{\text{c}} \Delta t, \forall k \in \mathcal{T}_{t} \label{energy_constraint}\\ 
  & \hat{\mathbf{u}}^{k} = \left[ \begin{array}{c}
      \mathbf{u}^{k}  \\
      -\mathbf{u}^{k}
 \end{array} \right], \quad \mathbf{x}^{k} =  \left[ \begin{array}{c}
      \mathbf{s}^{k}  \\
      \hat{\mathbf{u}}^{k} 
 \end{array} \right],  \forall k \in \mathcal{T}_{t}   \label{c_inputs}\\ 
  & \mathbf{u}^{k} \in \mathcal{U}^{n_{\text{z}}}, \forall k \in \mathcal{T}_{t}.  \label{c_domain}
\end{alignat} 
}
We consider a time-varying rates for energy $\lambda_{\text{E}}^{t} \geq 0 $, and a fixed rate on peak power demand $\lambda_{\text{P}} \geq 0 $, as utilities usually bill commercial consumers for both. Peak demand fees are based on the maximum power usage observed during a month~\cite{Rate_M}. Because of the lack of whole building's power profiles due, for example, to submetering, we aim to minimize the peak demand induced by the \glspl{rtu} over the prediction horizon as a proxy. Constraints~(\ref{dynamics})$-$(\ref{upper_temp_bound}) represent the thermal comfort zone, where $\bar{\boldsymbol{\theta}}^{t} \in \mathbb{R}^{n_{\text{z}}}$, \textcolor{blue}{$t\in\{1,2,\dots\}$}, is the upper bound on \gls{iat}, for all $n_{\text{z}}$ zones. The building's thermal dynamics are described by a parametrized \gls{icrnn}, $f_{\text{ICRNN}}:\mathbb{R}^{(2n_{\text{u}}n_{\text{z}}+n_{\text{s}}) \times T} \mapsto \mathbb{R}^{n_{\text{z}} \times T}$ that takes as inputs extended controls $\hat{\mathbf{u}}^{t} \in \mathcal{\hat{U}}$, and the $n_s \in \mathbb{N}$ building's state variables, i.e., exogenous variables, defined for all \textcolor{blue}{$t\in\{1,2,\dots\}$}, as $\mathbf{s}^{t}\in \mathbb{R}^{n_{\text{s}}}$. \textcolor{blue}{Details about the ICRNN, including its architecture and training procedure, are presented in Section~\ref{icrnn_train}}. We omit the lower bound on \gls{iat} $\underaccent{\bar}{\boldsymbol{\theta}}^{t} \in \mathbb{R}^{n_{\text{z}}}$,  \textcolor{blue}{$t\in\{1,2,\dots\}$}, to maintain convexity \textcolor{red}{as superlevel sets of~(\ref{dynamics}) are not garanteed to be convex~\cite{bunning_MPC}}. This relaxation is not limiting \textcolor{blue}{for cooling operations} because, by minimizing energy consumption and peak power, the \gls{iat} will tend to be close to the upper bound. \textcolor{blue}{We plan to extend this approach to heating operations in future work}. Constraint~(\ref{lock-out}) prevents equipment toggling, as short cycling significantly decreases the efficiency of units~\cite{cycling_RTU}. It ensures that each \gls{rtu} component remains in the same state during $\rho \in \mathbb{N}$ time steps when transitioning from \textsc{on} to \textsc{off}. \textcolor{blue}{Here, $\mathbf{0}\in \mathbb{R}^{n_{\text{z}}n_{\text{u}}}$ denotes the zero vector}. Constraint~(\ref{energy_constraint}) represents the energy consumed during a control round. Constraint~(\ref{c_inputs}) defines the inputs of the \gls{icrnn}, and~(\ref{c_domain}) states the domains of the optimization variables. \textcolor{blue}{Considering our \gls{icrnn} architecture further described in Section~\ref{icrnn_train}}, the problem~(\ref{obj})$-$(\ref{c_domain}) is a mixed-integer piecewise linear optimization problem. When the binary variables are relaxed to continuous ones, its objective function and its feasible set are convex, and the proposed \gls{mpc} is a convex problem~\cite{boyd}. Convexity of the relaxed optimization problem guarantees that the solver avoids getting stuck in local minima, and therefore promotes higher solution quality with respect to conventional \gls{nn}-based \gls{mpc}.

\subsubsection{Demand Bidding Strategy}\label{demand_bidding_modelling}

We now propose a strategy that allows \gls{rtu}-based \glspl{cb} participation in a \gls{dr} program, \textcolor{blue}{leveraging an approach that balances modelling accuracy with computational efficiency. Our aim is to enhance bid accuracy and improve the estimation of potential gains.} Participants can bid capacity in the real-time market, which operates on 5-minute intervals, matching the \gls{cb}'s operating timescale. Market parameters such as the minimum bid size, the trading intervals time span and gate closure times for bid submission tend to differ in each market~\cite{market_overview}. \textcolor{blue}{We formulate a model that can be easily adapted to different intraday markets like CAISO, AEMO, and PJM.} We draw inspiration from the \gls{fcas} market of the \gls{nem} requirements for the response speed and duration. \gls{fcas} are market-based ancillary services that are used to maintain the grid's frequency to a normal operating point by balancing generation and demand. Frequency control services are further divided into regulation and contingency services. Here, we target contingency delayed-raise and delayed-lower services which require participants to provide an increase or a curtailment of power and maintain it for 5 minutes. When called, participants must respond within 5 minutes. Participants are rewarded based on their availability, regardless of if they are called or not~\cite{aemo_2021}. Finally, we make the following assumptions regarding some of the market requirements.
\begin{enumerate}
\item The service is rewarded based on deviations from a baseline of \gls{hvac} power data provided by the \gls{iso}. 
\item We adopt a price-taker bidding strategy. This assumption is not too restrictive, as markets like CAISO only allow \gls{dr} participants to bid as price-takers in the real-time market~\cite{caiso_bid}. 
\item We assume that bids in the day-ahead market have already been submitted if it is a necessary step to participate in the intraday market.  
\item We omit the minimum-sized bid requirement. If necessary, multiple individual bids can be aggregated. %
\end{enumerate} 


\textcolor{blue}{We now model common market operations to ensure the adaptability and transferability of our approach. We consider the following timeline of events for a given trading interval starting at $t+3$}. Participants submit their bids to the \gls{iso} before the market's gate closure at market interval $t$. Once closed, bid modifications are no longer possible. Then follows the settlement period, during which the \gls{iso} receives all bids and secures enough resources. By the end of this period, participants receive a guarantee of payment if their bid is cleared in the market at $t+1$. Subsequently follows a standby period where participants are waiting to receive the \gls{dr} signal $\sigma_{-}^{t+3} \in \{0,1\}$ from the \gls{iso}, where $\sigma_{-}^{t+3}$ represent being called ($\sigma_{-}=1$) or not ($\sigma_{-}=0$). Upon its reception at $t+2$, participants are \textcolor{blue}{required} to fulfill their service bid. 

Next, we formulate a real-time bidding strategy for a single \gls{cb}. We define $\mathcal{W}_{t}\subseteq\mathcal{T}_{t}$ as the set of market intervals closed for bid modifications. Let $ \lambda_{-}^{t} \geq 0$, \textcolor{blue}{$t\in\{1,2,\dots\}$}, be the \glspl{mcp} for energy curtailment. Let $c_{-}^{t} \geq 0$, \textcolor{blue}{$t\in\{1,2,\dots\}$}, be the amount of capacity previously settled by the \gls{iso}. Let $p^{t}_{\text{baseline}} \geq 0$, \textcolor{blue}{$t\in\{1,2,\dots\}$} be the \gls{hvac} power baseline. The \gls{mpc} for demand bidding is provided in~(\ref{DR_obj})$-$(\ref{reward}) and discussed next.
\begin{align}
\underset{\hat{\mathbf{u}}^{k}, k\in\mathcal{T}_{t}}{\text{min}} &  \sum_{k \in \mathcal{T}_{t}} \textcolor{blue}{\lambda^{k}_{\text{E}}}E^{k}-\sum_{ k \in \mathcal{T}_{t}\setminus \mathcal{W}_{t}}\lambda_{-}^{k}\Delta E^{k}_{-} - \sum_{k \in \mathcal{W}_{t}}\lambda^{k}_{-}R_{-}^{k}\label{DR_obj} \\  
 \text{s.t. } \quad  & (\ref{dynamics})-(\ref{c_domain}) \nonumber  \\ 
& \Delta E_{-}^{k}= \left(p^{k}_{\text{baseline}}- {\mathbf{u}^{k}}^\top  \mathbf{p}^{\text{c}}\right)\Delta t,\text{ }\forall k \in \mathcal{T}_{t}\setminus \mathcal{W}_{t} \label{adjusment} \\   
& p^{k}_{\text{baseline}} - {\mathbf{u}^{k}}^\top  \mathbf{p}^{\text{c}}  \geq c^{k}_{-}, \text{  }\text{if }  \sigma^{k}_{-}=1,  \forall k \in \mathcal{W}_{t}\label{bid_down}  \\
& R_{-}^{k} = c_{-}^{k}\sigma^{k}_{-}, \text{ } \forall k \in \mathcal{W}_{t}\label{reward}
\end{align}   
The objective function~(\ref{DR_obj}) is defined in terms of the energy cost, a penalty component, and rewards for curtailment services. When the power consumption exceeds the baseline, the penalty term is positive and is proportional to the deviation from the baseline. Conversely, when energy consumption is below the baseline, the penalty term is negative and equal to the reward for the services provided. This formulation is chosen to ensure the problem's convexity and to eliminate the need to introduce equality constraints or supplementary variables. This consideration is especially relevant as the problem is solved through \gls{dfo}, which will be discussed in Section~\ref{case_study}. Constraint~(\ref{adjusment}) represents the amount of energy adjustments made with respect to the baseline. Once a bid is cleared in the market, participants are expected to provide the adequate service when called by the \gls{iso}. This is modelled by constraint~(\ref{bid_down}). When solving the \gls{mpc} problem, we assume $\sigma_{-}^{t+3}=1$ from the market closure until the end of the bid settlement period. After this period, it is updated to its true value. If our bid is cleared in the market, it remains set to~1 during the standby period until the reception of the true dispatch signal.  Finally, constraint~(\ref{reward}) defines the rewards. 





\section{Numerical Case Studies} \label{case_study}
In this section, we conduct numerical experiments on a model based on the 2-zone building from the Modelica Buildings Library~\cite{WetterZuoNouiduiPang2014}, and use Dymola as the compiler. The library supports multi-zone heat transfer and multi-zone airflow~\cite{WetterZuoNouiduiPang2014}. \textcolor{blue}{The power ratings of the RTU components are $p_{\text{c}1}=p_{\text{c}2}=2.272$~kW and $p_{\text{f}}=0.637$~kW.} \textcolor{blue}{The simulated experiments are taking place in Miami from January~1\textsuperscript{st}--4\textsuperscript{th},~2021~\cite{energyplusweatherdata}, a period requiring cooling}. \textcolor{blue}{Given that the control round of each \gls{rnn}-based MPC has a duration of 5 minutes, we have limited the experiment duration. However, we present results across various \gls{dr} settings over full days of operations, considering different comfort zones based on occupancy.} Specifically, the thermal comfort zone of \gls{iat} is set to [20$^{\circ}$C, 24$^{\circ}$C] during occupancy hours, and to [18$^{\circ}$C, 28$^{\circ}$C] during unoccupied periods. The simulation environment is employed for data generation for model training, and subsequently for assessing the performance of the controllers. Model accuracy is presented in Section \ref{model_training}, and then the controller performance is discussed in Section~\ref{Controller_performances}.
\subsection{Modelling Accuracy} \label{model_training} 

\begin{figure}[tb]
\centering
\subfloat[ICRNN model in zone 2.]{\includegraphics[trim=0.45cm 0.5cm 0.45cm 0.45cm, clip,width=43mm]{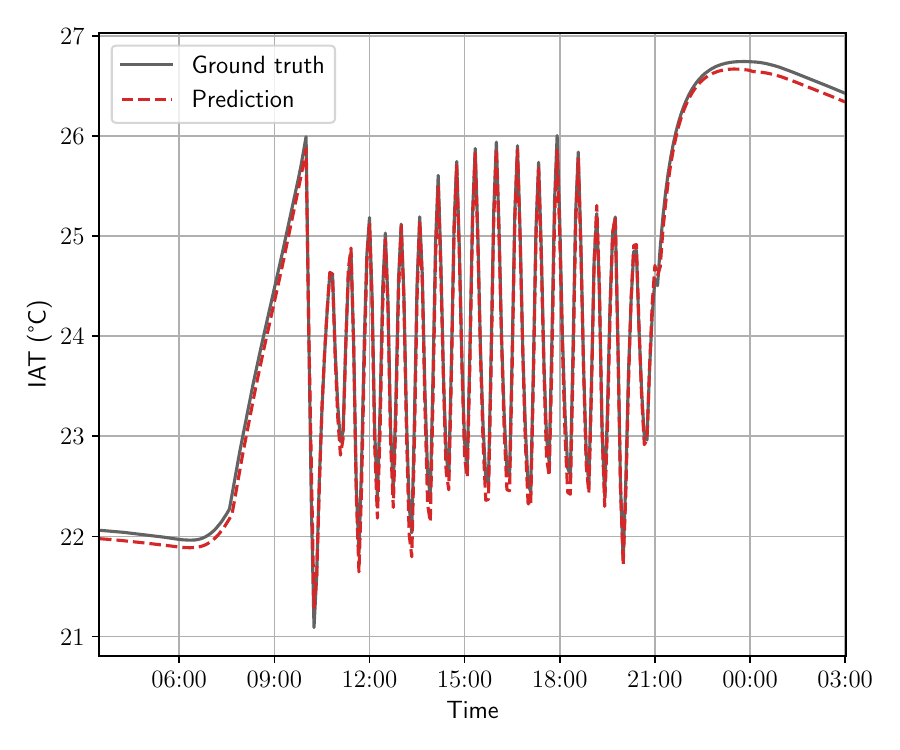}%
\label{subfig:Convex_zone2}} \\
\subfloat[Linear model in zone 2.]{\includegraphics[trim=0.45cm 0.5cm 0.45cm 0.45cm, clip,width=43mm]
{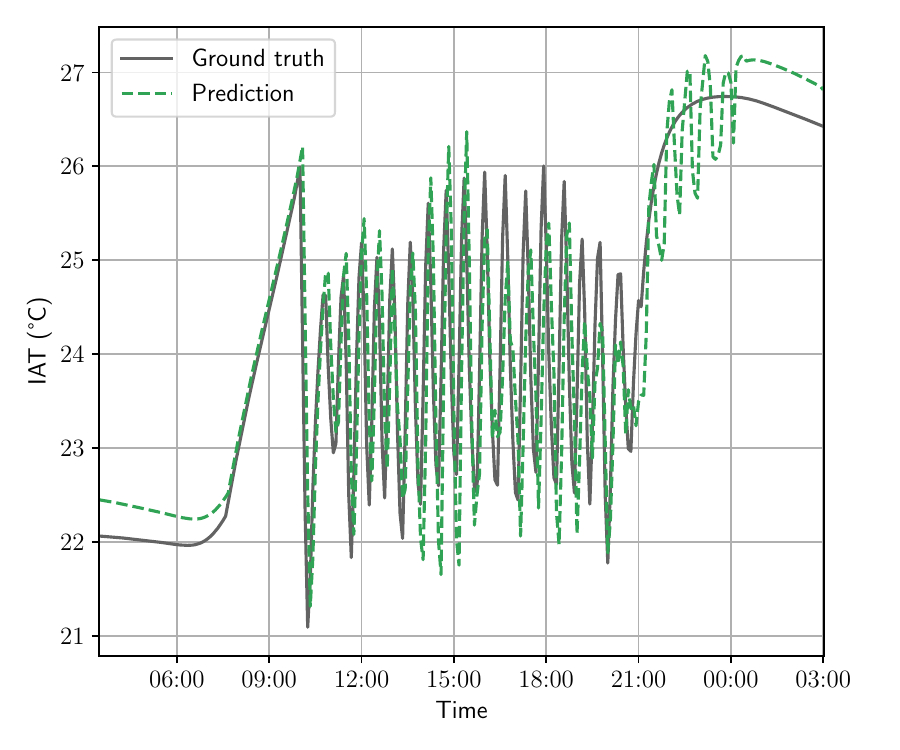}%
\label{fig:Day3_z2}}
\subfloat[LSTM model zone 2.]{\includegraphics[trim=0.45cm 0.5cm 0.45cm 0.45cm, clip,width=43mm]
{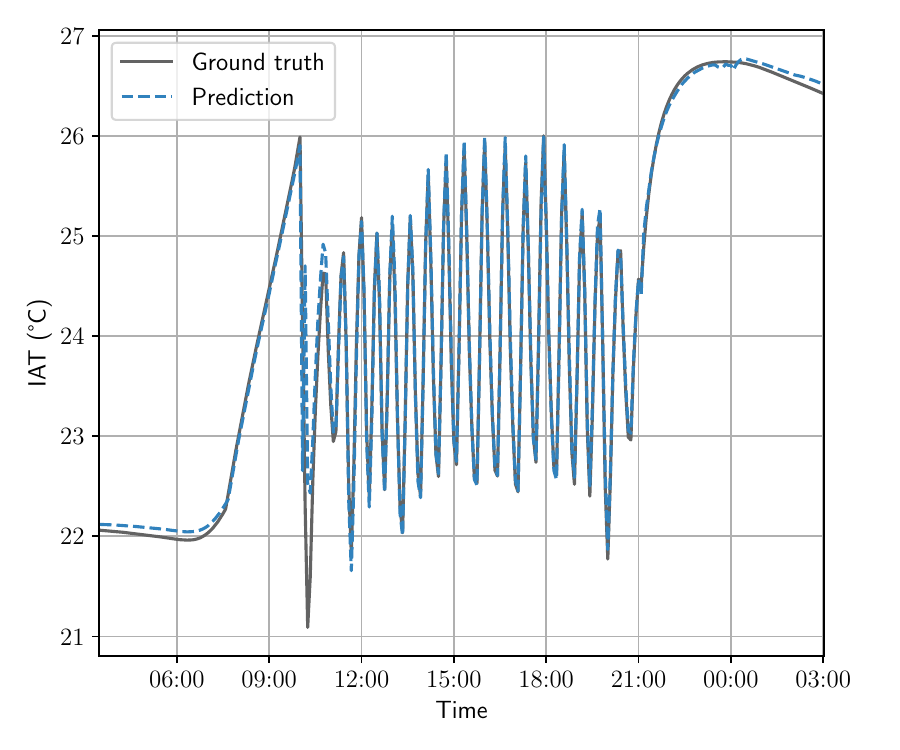}%
\label{subfig:lstm_zone2}}
\caption{Thermal dynamics fitting comparison.}
\label{fig:fit}
\end{figure}

We now describe the procedure to train the \gls{icrnn} and its benchmarks. Results are presented in Section~\ref{results_accuracy}.
\subsubsection{\gls{icrnn}}\label{icrnn_train}Our inference approach for multi-step-ahead prediction is based on a multi-output sequence-to-sequence model with a prediction horizon of 2h and a past data horizon of 3h. We use 6 months of data with a 5-minute sampling interval. The dataset is randomly sampled to form training, validation, and testing sets, each comprising 60\%, 20\%, and 20\% of the full dataset size, respectively. Following this, we apply min-max scaling to the data. For feature selection, a correlation analysis is used to target the relevant features and then a forward selection method is applied. The features selected include the extended controls $\mathbf{\hat{u}}^{t}$ and the state variables $\mathbf{s}^{t}$. The state variables consist of the \gls{oat}, the \gls{ghi}, \textcolor{blue}{the time of the day and of the week} in sin/cos encoding used as representations of the building occupancy. As for the target variables, the \gls{icrnn} is used to predict $\Delta \boldsymbol{\theta}^{t}$~\cite{bunning_MPC}. Predicting $\Delta \boldsymbol{\theta}^{t}$ instead of the \gls{iat} leads, in our case, to a higher modelling accuracy. We use a ReLU activation function for $\phi_{1}$ and a linear activation function of the output layer $\phi_{2}$. Hyperparameter tuning is performed with \texttt{Oríon}, a black-box optimization framework~\cite{Orion}. The tuned hyperparameters include the number of hidden units, the number of hidden layers, and the learning rate. Finally, early stopping is implemented to prevent overfitting during training.
\subsubsection{Benchmark} To further investigate the trade-off between modelling accuracy and computational tractability, we use a linear model and \textcolor{blue}{a \gls{lstm} \gls{rnn}} as benchmarks. In this work, the nonconvex \gls{mpc} serves as a benchmark for model accuracy, but is computationally intractable. The linear \gls{mpc} acts as a benchmark for computational tractability but with limited modelling abilities.  Lastly, the greedy controller represents control simplicity.
 
\paragraph{Linear} \textcolor{red}{We use a multi-output first-order \gls{arx} model similar to~\cite{first_order_arx}} and employ a recursive approach for generating multi-step-ahead prediction. \textcolor{blue}{Input features include OAT, time of day of week in sin/cos encoding at $t$,  GHI from $t-3$ to $t$, and $\Delta \boldsymbol{\theta}^{t}$ from $t-4$ to $t-1$.}
\paragraph{Nonconvex} We use an \gls{lstm} \gls{nn}. \glspl{lstm} have been widely applied to building modelling because of their high accuracy~\cite{su2022}. However, their highly \textcolor{blue}{nonlinear} structure can potentially lead the solver to converge to a local minimum. We follow the same process for feature selection, training, and hyperparameter tuning as we do for the \gls{icrnn} model. 
\paragraph{Greedy} We employ an \gls{sp} controller that takes decisions based on the current occupancy status and temperature conditions relative to comfort bounds and dead bands. 

\subsubsection{Results}\label{results_accuracy} 
We compare our modelling approach with benchmark models. Figure~\ref{fig:fit} illustrates the prediction performance over a 24-hour period in one of the two zones. To fit the \gls{nn} models, we employ a recurrent layer of 60 units for the \gls{icrnn} and 45 units for the \gls{lstm}. We compute the corresponding \gls{rmse} and their standard \textcolor{blue}{deviation} for each predictive model. In terms of fitting performances, the linear model produces the largest error values with a \gls{rmse} of ($0.45 \pm 0.04$)$^{\circ}$C. \textcolor{red}{This is primarily due to error propagation over time resulting from the recursive approach and its limitations in capturing nonlinear patterns such as coupling between zones~\cite{MORK2022}}. The \gls{lstm} model, conversely, shows the lowest error values with a \gls{rmse} of ($0.09 \pm 0.01$)$^{\circ}$C. An \gls{lstm} \gls{nn} has a memory cell that can capture information from earlier time steps and store it for later use, which helps addressing the problem of vanishing or exploding gradient~\cite{FAN_2019}, thereby contributing to a more accurate representation of thermal inertia. Our \gls{icrnn} model shows significant improvements when compared to the linear model with a \gls{rmse} of ($0.14 \pm 0.03$)$^{\circ}$C. Because the \gls{icrnn} is a piecewise linear convex approximation, it has representation limitations. It remains a competitive alternative to the \gls{lstm}. 


\subsection{Controller Performance}\label{Controller_performances}

\begin{table}[tb]
    \centering
    \caption{Controller comparison for the flat-rate pricing.}
    \begin{tabular}{lccc}
        \Xhline{0.75pt} \textbf{Metrics} &  \textbf{Day 1}  &  \textbf{Day 2} &  \textbf{Day 3}  \\
       \hline 
       \textbf{Greedy } &  & & \\
         \hline Avg. discomfort [$^{\circ}$C] & 0.307 & 0.360 &  0.437 \\
       Energy [kWh] & 52.39 & 52.73 & 50.32  \\ 
       Toggling & 0 & 0 & 0 \\ 
          \hline 
       \textbf{Linear} &  & & \\
         \hline Avg. discomfort [$^{\circ}$C] & 0.015 & 0.014 & 0.052  \\
       Energy [kWh] & 95.03 & 93.40 & 94.16  \\ 
       Toggling & 0 & 0 & 0 \\ 
       \hline 
       \textbf{Convex} &  & & \\
       \hline Avg. discomfort [$^{\circ}$C] & 0.069 & 0.005  & 0.066  \\
       Energy [kWh] & 52.63 & 60.12 & 30.32 \\ 
       Toggling & 0 & 0 & 0  \\ 
          \hline 
       \textbf{Nonconvex} &  & & \\
         \hline Avg. discomfort [$^{\circ}$C] & 0.003 & 0.005 & 0.006  \\
       Energy [kWh] & 59.98 & 67.11 & 62.37  \\ 
       Toggling & 0 & 0 & 0  \\ 
       \Xhline{0.75pt}
    \end{tabular}
    \label{tab:flat_tarif}
\end{table}

We now assess the performance of the controllers under different pricing mechanisms and \gls{dr} programs. 

In our control experiments, we introduce Gaussian noise to simulate uncertainty in time-varying data and compute standard deviations from historical data. Specifically, we apply noise to the \gls{oat} and the \gls{ghi}. To ensure a reasonable variation over a 2-hour horizon, the standard deviations are scaled down by a factor of 10 for \gls{iat} and 50 for \gls{ghi}, and the amount of noise is increased linearly across the prediction horizon. In Section \ref{demand_bidding}, the same process is applied to introduce noise in energy prices and \glspl{mcp}, with standard deviations adjusted by a factor of 10. As for the performance metrics, we evaluate thermal discomfort, defined at time step $t$ as $\mathbf{1}^{\top}\left([\underaccent{\bar}{\boldsymbol{\theta}}^{t}-\boldsymbol{\theta}^{t}]^{+}+ [\boldsymbol{\theta}^{t}-\bar{\boldsymbol{\theta}}^{t}]^{+}\right)\Delta t$, where $\mathbf{1} \in \mathbb{R}^{n_{z}}$ is the one vector, and toggling based on the satisfaction of constraint~(\ref{lock-out}).

In terms of resolution methods, the \gls{mpc} based on the linear model is solved \textcolor{blue}{with \texttt{MOSEK}~\cite{mosek}, which returns the global optimum}. In case of infeasibility, thermal comfort bounds are relaxed. The \gls{nn}-based \gls{mpc}s are solved with the blackbox optimization software \texttt{NOMAD}~\cite{Nomad3} using the \texttt{Mesh Adaptive
Direct Search}~\cite{MADS_2006} algorithm, which allows for discrete-\gls{dfo}\textcolor{blue}{~\cite{audet2019mesh}}. For the nonconvex \gls{mpc} specifically, a \texttt{Variable Neighborhood Search}~\cite{VNS} metaheuristic is added to try to escape local minima. In this optimization framework, all constraints, i.e., constraints (\ref{dynamics})$-$(\ref{lock-out}) and~(\ref{bid_down}) are \textcolor{blue}{handled by the} \texttt{Progressive Barrier} approach, which allows constraint violations. In case of infeasibility, \texttt{NOMAD} yields the solution that minimizes the sum of \textcolor{blue}{squares} of the violations. To initiate the optimization process, the solver requires an initial point. This enables us to leverage the previous solution in a rolling horizon strategy, to effectively guide the new search. To reduce the search space, binary controls presented in Section~\ref{system_description} are converted into integer variables. 

\subsubsection{Flat-Rate Pricing}\label{Flat-rate_tariff}


\begin{figure*}[!t]
\centering
\subfloat[Zone 1 on day 2.]{\includegraphics[trim=0.25cm 1.cm 1cm 0.5cm, clip,width=\columnwidth]{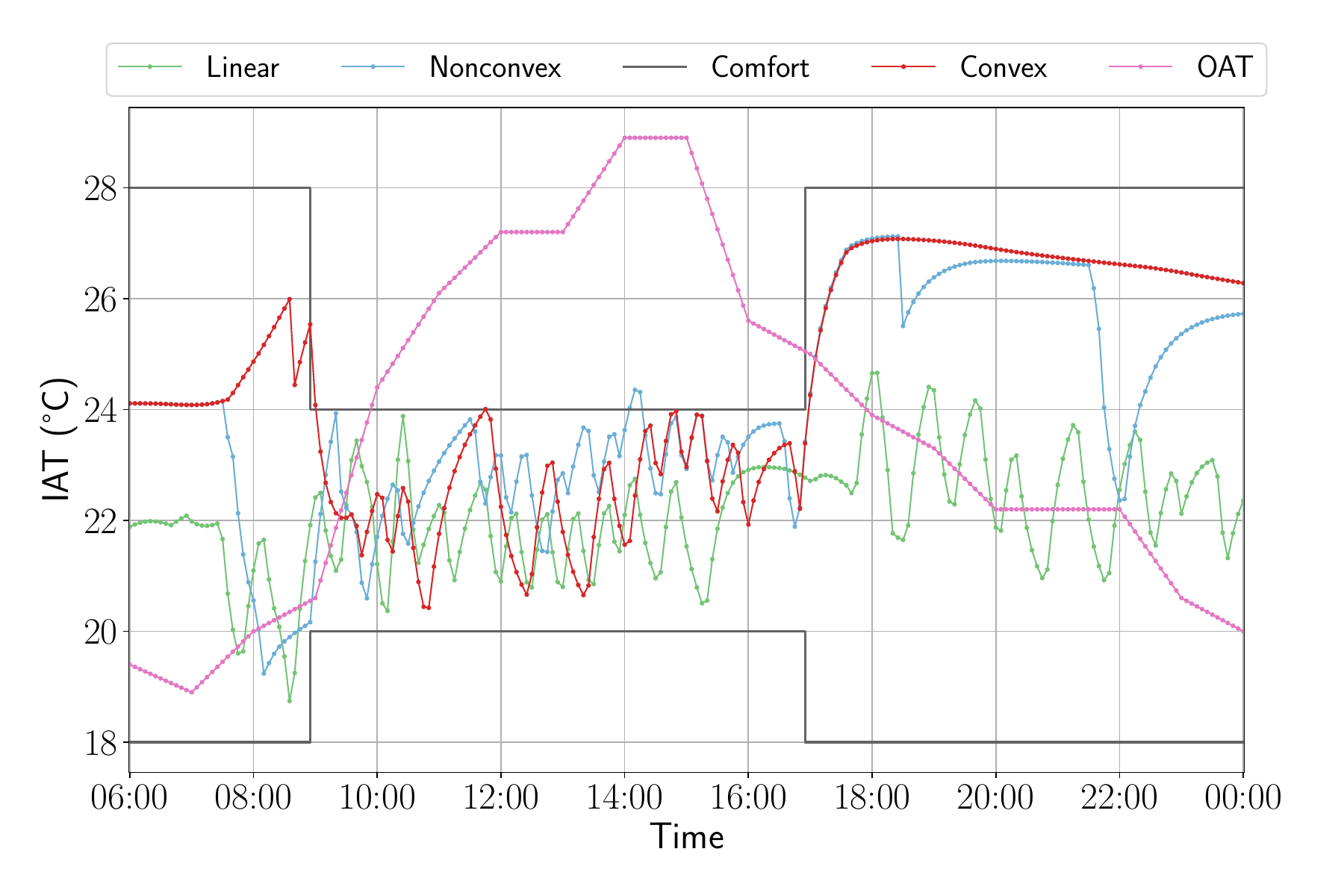}%
\label{fig:Day2_z1}}
\subfloat[Zone 2 on day 3.]{\includegraphics[trim=0.25cm 1.cm 1cm 0.5cm, clip,width=\columnwidth]{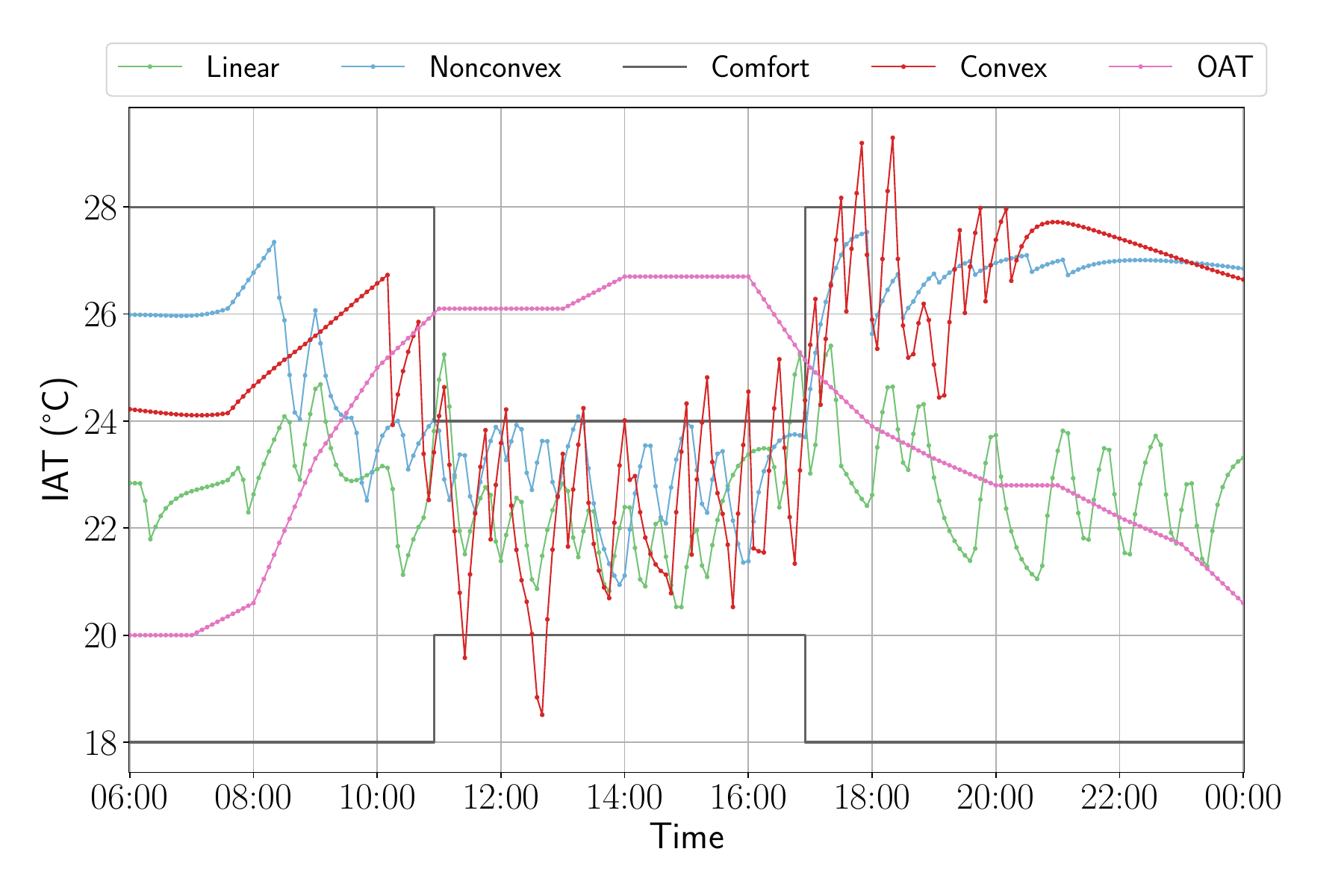}%
\label{fig:Day3_z2}}
\caption{Controller comparison: \gls{iat} profiles during 2 simulated days under a flat-rate pricing.}
\label{fig:IAT_ts}
\end{figure*}

We conducted a 3-day experiment implementing the \gls{mpc} problem described in~(\ref{obj})$-$(\ref{c_domain}), considering both time-invariant energy and power pricing. The energy rate is set to \$0.05303/kWh, and the peak power rate to \$14.58/kW, similarly to~\cite{Rate_M}. Results for all zones combined are presented in Table~\ref{tab:flat_tarif}.


In a cooling experiment, the greedy controller toggles around the upper bound on thermal comfort. It thus acts as a more stringent benchmark for energy consumption but induces a significant amount of thermal discomfort. As shown in Table~\ref{tab:flat_tarif}, on both day 1 and day 3, the greedy controller consumes either the same amount or more energy than the convex \gls{mpc}, while also causing significant thermal discomfort. These results highlight the importance of operating \gls{rtu}-\gls{hvac} efficiently. With the linear \gls{mpc}, the controller tends to inefficiently overcool the system, leading to a high energy consumption. This underscores the crucial role of model design in control methods like \gls{mpc}, as an inadequate model will lead to poor controller performance. Considering the convex approach, the \gls{icrnn} has a higher prediction error than the \gls{lstm}. In control, prediction errors may lead the \gls{iat} outside the thermal comfort zone, as presented in Figure~\ref{fig:IAT_ts}. Although the convex \gls{mpc} problem has only an upper bound on thermal comfort, the \gls{iat} is kept within the comfort zones \textcolor{blue}{most of the time}, as shown in Figure~\ref{fig:IAT_ts}\subref{fig:Day3_z2}.  In all instances, all controllers avoid toggling of equipment. However, none of them manage to reduce the power peak consumption of 10.37~kW. Given the small number of zones and the experiments taking place during hot days, it may not be feasible to achieve such reduction. Overall, when compared to the more commonly implemented control method, viz., the greedy controller, the convex \gls{mpc} is the only approach that reduces both energy consumption and discomfort, with respective averages of 8.42\% and 86.95\% across all simulation days.


\subsubsection{Demand Bidding} \label{demand_bidding}
\begin{table}[tb]
    \centering
     \caption{Controller comparison for \gls{dr} programs.}
     \resizebox{\columnwidth}{!}{
    \begin{tabular}{lccccc}
    \Xhline{0.75pt} 
    \multicolumn{6}{c}{\textbf{Demand bidding}} \\
     \Xhline{0.5pt} \multirow{2}{*}{\begin{minipage}[t]{0.75cm} \textbf{Controller} \end{minipage}} & \multirow{2}{*}{\begin{minipage}[t]{1.cm} \textbf{\textbf{Energy} \textbf{[kWh]}}\end{minipage}} & \multirow{2}{*}{\begin{minipage}[t]{1.6cm} \textbf{Avg. dis-\\comfort} \textbf{[$^{\circ}$C]}\end{minipage}}  & \multirow{2}{*}{\begin{minipage}[t]{1.cm} \textbf{Toggling}\end{minipage}} & \multirow{2}{*}{\begin{minipage}[t]{1.cm} \textbf{Net cost} \textbf{[\$]} \end{minipage}} & \multirow{2}{*}{\begin{minipage}[t]{1.cm} \textbf{Savings} \textbf{[\%]} \end{minipage}} \\
        &  &   &   &  &   \\
     \hline Greedy & 52.39 & 0.307 & 0 & 1.28 & -\\
        Linear & 82.55 & 0.010 & 0 & 1.76 & 0.76\\
        Convex & 45.62 & 0.092  & 2 & 0.95 & 22.02\\
        Nonconvex & 52.44 & 0.029 & 2  & 1.12 & 17.07\\
    \Xhline{0.5pt}
     \multicolumn{6}{c}{\textbf{Time-of-use}}\\
    \Xhline{0.5pt} \multirow{2}{*}{\begin{minipage}[t]{0.75cm} \textbf{Controller}\end{minipage}} & \multirow{2}{*}{\begin{minipage}[t]{1.cm} \textbf{\textbf{Energy} \textbf{[kWh]}}\end{minipage}} & \multirow{2}{*}{\begin{minipage}[t]{1.6cm} \textbf{Avg. dis-\\comfort} \textbf{[$^{\circ}$C]}\end{minipage}}  & \multirow{2}{*}{\begin{minipage}[t]{1.cm} \textbf{Toggling}\end{minipage}} & \multirow{2}{*}{\begin{minipage}[t]{1.cm} \textbf{Net cost} \textbf{[\$]} \end{minipage}} &\\
       &  &  &    &  &\\
     \hline Greedy  & 52.39 & 0.307 & 0.0 & 7.82 &  -\\
     Linear & 93.14 & 0.016 & 0.0 & 12.69 & -\\
    Convex & 52.29 & 0.033 & 1.0 & 8.115 & - \\
     Nonconvex & 61.49  & 0.000 & 0.0 & 9.53 & -\\
     \Xhline{0.5pt}
     \multicolumn{6}{c}{\textbf{Critical peak rebate}}\\
     \Xhline{0.5pt} \multirow{2}{*}{\begin{minipage}[t]{0.75cm} \textbf{Controller}\end{minipage}} & \multirow{2}{*}{\begin{minipage}[t]{1.cm} \textbf{\textbf{Energy} \textbf{[kWh]}}\end{minipage}} & \multirow{2}{*}{\begin{minipage}[t]{1.6cm} \textbf{Avg. dis-\\comfort} \textbf{[$^{\circ}$C]}\end{minipage}}  & \multirow{2}{*}{\begin{minipage}[t]{1.cm} \textbf{Toggling}\end{minipage}} & \multirow{2}{*}{\begin{minipage}[t]{1.cm} \textbf{Net cost} \textbf{[\$]} \end{minipage}} & \multirow{2}{*}{\begin{minipage}[t]{1.cm} \textbf{Savings} \textbf{[kWh]} \end{minipage}} \\
      &  &  &  & &  \\  
     \hline Greedy  & 52.39 & 0.309 & 0 & -0.31 & 7.80 \\ 
     Linear &  86.79  &   0.060 & 0 &  3.53 &  5.57 \\
    Convex & 51.70 & 0.054 & 0 & -0.10 & 7.33\\ 
     Nonconvex & 60.04 & 0.002 & 1 & 3.18 &  2.51\\ 
     \Xhline{0.75pt}
    \end{tabular}}
    \label{tab:DR}
\end{table}

\begin{figure}[tb]
    \centering
    \subfloat[Linear \gls{mpc}.]{\includegraphics[trim=0.25cm 1.cm 1cm 1cm, clip, width=82mm]{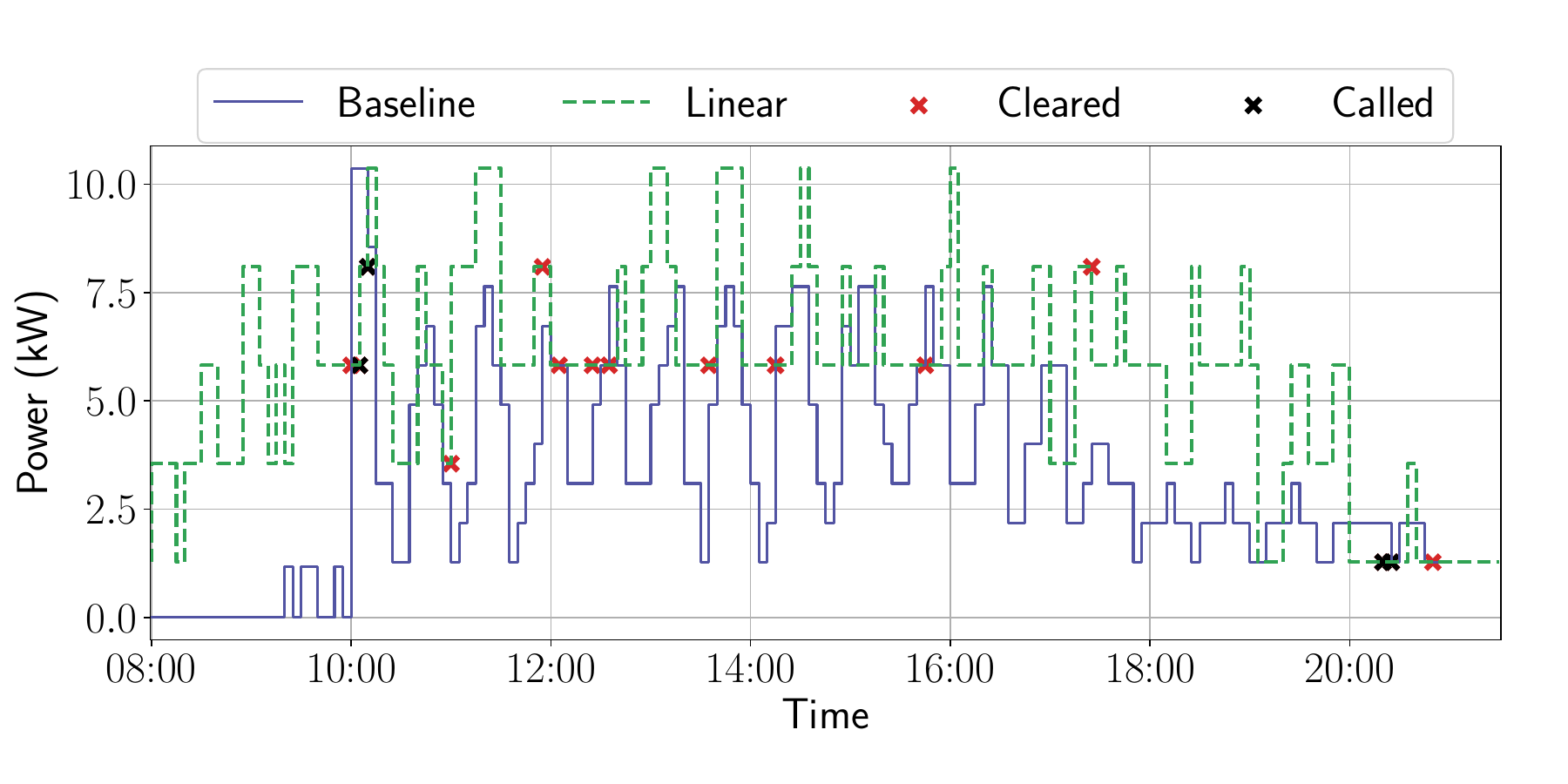}\label{fig:bid_linear}}\\
    \subfloat[Convex \gls{mpc}.]{\includegraphics[trim=0.25cm 1.cm 1cm 1cm, clip, width=82mm]{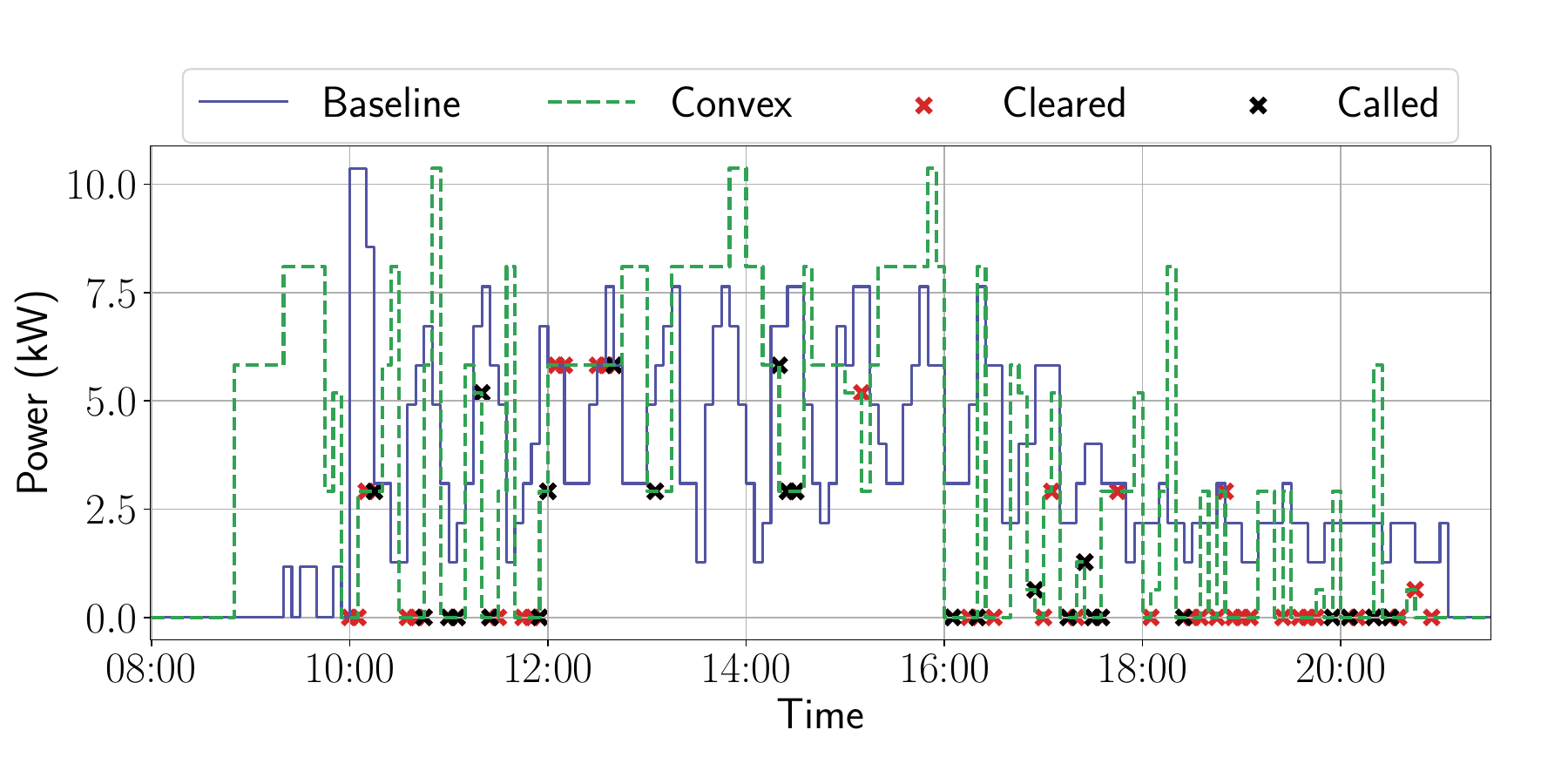}\label{fig:bid_icrnn}}\\
    \subfloat[Nonconvex \gls{mpc}.]{\includegraphics[trim=0.25cm 1.cm 1cm 1cm, clip, width=82mm]{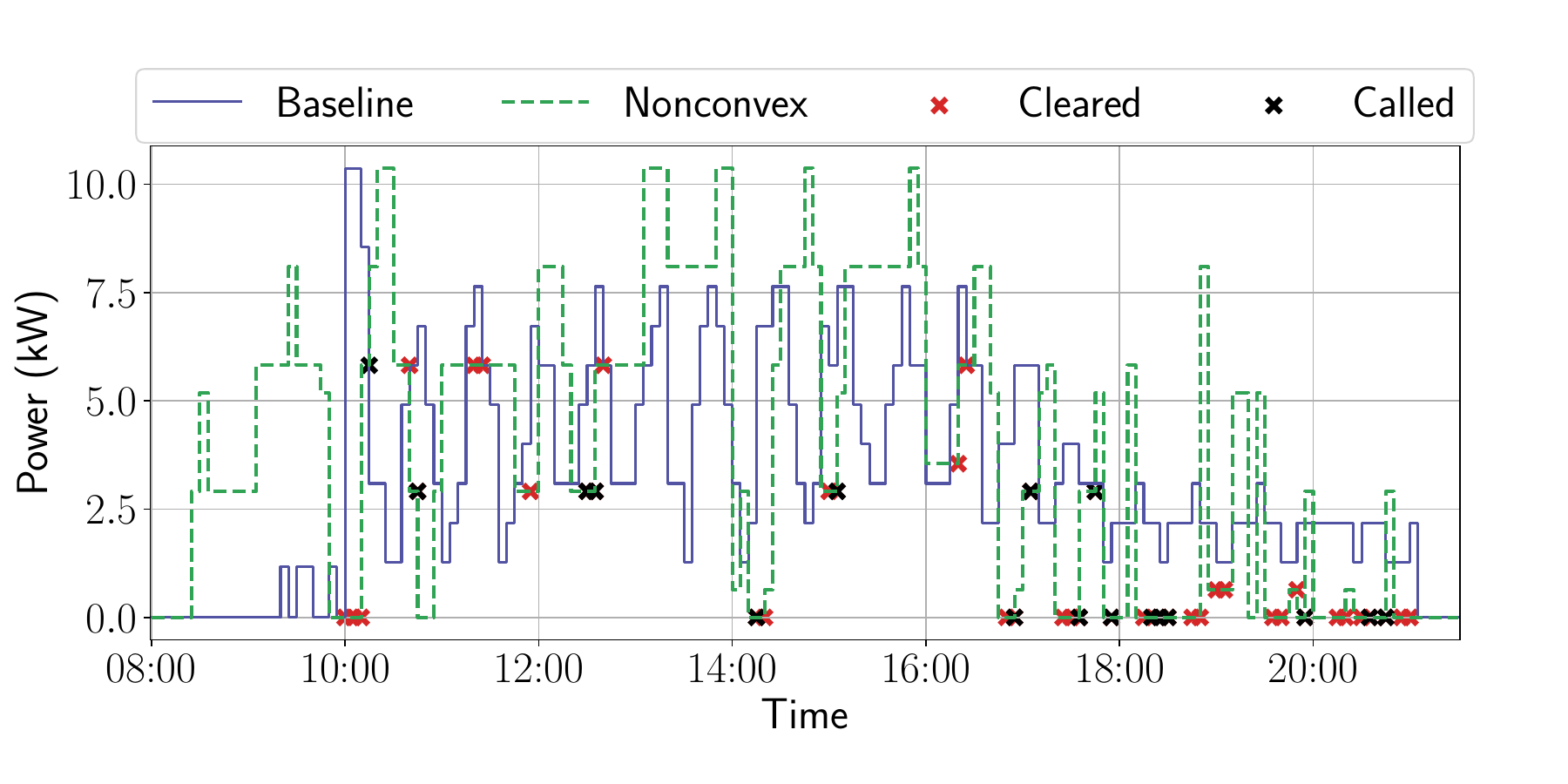}\label{fig:bid_nonconvex}}\label{subfig:bid_nonconvex}
    \caption{Power profiles for the demand bidding program.}
    \label{fig:bid}
\end{figure}


We now implement the demand bidding strategy presented in Section~\ref{demand_bidding_modelling}. In this experiment, a common baseline consumption is set for all \glspl{mpc}, and is established by averaging historical data of the corresponding day of the week. \textcolor{blue}{The market's gate closure time is set to 15 minutes before the start of a trading interval.} The \gls{iso}'s decision $\sigma^{t}_{-}$, \textcolor{blue}{$t\in\{1,2,\dots\}$}, on whether to call a service or not is modelled by independent and identically distributed Bernoulli random variables. We assume a probability of~0.9 of being cleared in the market, as we participate as a price-taker, and a probability of being called of~0.4. Real-time energy prices and \glspl{mcp} are retrieved from~\cite{PJM_LMP, PJM_MCP}. Savings from participation are calculated with respect to a benchmark for each thermal model, obtained by solving the control problem defined by~(\ref{dynamics})$-$(\ref{DR_obj}) with $\lambda^{t}_{-}=0$  $\forall t$, \textcolor{blue}{and real-time energy prices are used for $\lambda^{t}_{E}$.} 

In this experiment, the \gls{cb} participates in the real-time market by submitting bids over a 24-hour period. Detailed results for all zones combined are presented in Table~\ref{tab:DR}. The convex \gls{mpc} attains the lowest energy consumption and net cost among all controllers. As for the average thermal discomfort, the proposed approach achieves a reduction of 70.03\% compared to the greedy approach. Both the convex and nonconvex \glspl{mpc} show a comparable amount of toggling. In general, a tolerable level of cycling is acceptable, and this remains within reasonable limits, as the results presented include all components of the two \glspl{rtu}. Regarding savings, the convex approach achieves the largest amount of savings. This can be attributed to the fact that the convex \gls{mpc} submits significantly more bids into the market, as presented in Figure~\ref{fig:bid}. The convex approach bids 12.35~kWh through 66 individual bids, and is called 28 times. In contrast, the linear \gls{mpc} submits 2.72~kWh through 16 bids, and the nonconvex submits 7.07~kWh through 47 bids. Finally, the convex approach ends up providing more flexibility to the grid, delivering 4.13~kWh of service, compared to 2.49~kWh for the nonconvex and 1.33~kWh for the linear. During the experiment, all controllers successfully fulfilled their bids. In a real market context, participants who fail to do so could incur a substantial penalty. If feasibility is a concern, constraint~(\ref{bid_down}) could be treated with \textcolor{blue}{this} \texttt{Extreme Barrier}~\cite{MADS_2006} approach to forbid violations.



\subsubsection{Time-of-Use}\label{Time_of_use} 


We now consider a \gls{tou} program, where rates and schedules are based on~\cite{TOU_ontario} for the summer period.

Results for a 24-hour experiment are presented in Table~\ref{tab:DR}, and include all zones combined. In this experiment, the convex controller attains the lowest energy consumption among all controllers. While it ranks second in terms of net cost, with a net cost 3.77\% higher than the greedy controller, it mitigates discomfort by 89.25\% with an equivalent energy usage.  

\subsubsection{Critical Peak Rebate}\label{Critical_peak_rebate}
We now consider a \gls{cpr} program.  \gls{cpr} is a program in which participants receive discounts on their electricity bills in exchange for reducing their demand during critical peak periods. The program utilizes a participant baseline to determine the payment settlement. From the consumer's perspective, \gls{cpr} is a voluntary and self-controlled initiative that incurs no penalties and requires minimal equipment~\cite{CPR_2021}. 

In this experiment, the \gls{mpc} problem is defined by the set (\ref{dynamics})$-$(\ref{c_domain}), and (\ref{adjusment}), with objective~(\ref{DR_obj}) where the reward term is not included. The electricity rate is structured with a base rate of \$0.076/kWh and a reward of \$0.55/kWh for each kWh reduced from the participant's baseline~\cite{Flex_D}. The baseline is defined as in Section~\ref{demand_bidding}. We consider events that take place during the summer, where peak demand typically occurs within the 12:00 PM to 6:00 PM window. To prepare for the event, program participants are assumed to be notified at least two hours in advance. 

In Table~\ref{tab:DR}, the results of all controllers are compared. The convex approach achieves the lowest energy consumption compared to other controllers without leading to equipment toggling. In terms of net cost, when compared to the greedy controller, it incurs an additional cost of \$0.21, but achieves a reduction of 82.52\% in terms of discomfort. In terms of energy savings, the convex approach is able to curtail 7.33~kWh from the baseline, which is only 6.02\% less than the greedy controller.

\subsection{Solution Quality} 
\begin{figure}[b]
\centering     
\subfloat[5 vs. 10 min.]{\includegraphics[trim=0.75cm 0.75cm 0.75cm 0.75cm, clip, width=43mm]{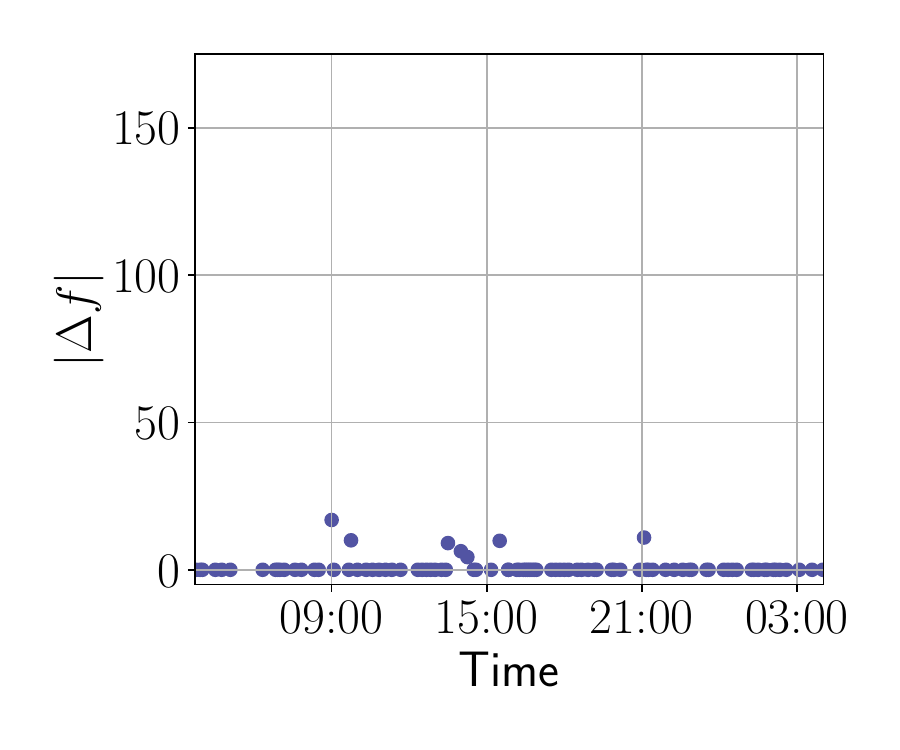}} 
\hfill
\subfloat[5 vs. 30 min.]{\includegraphics[trim=0.75cm 0.75cm 0.75cm 0.75cm, clip,width=43mm]{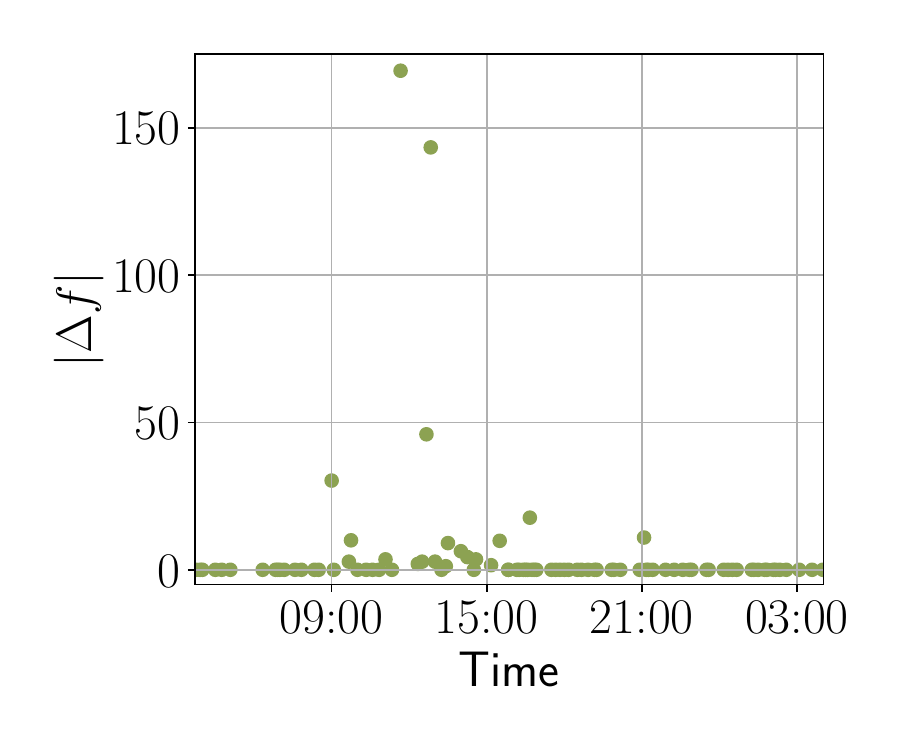}} 
\caption{Comparison of objective \textcolor{blue}{function} values based on computation time.}
\label{fig:optimiality_test}
\end{figure}

We now assess the quality of the solution obtained during a 5-minute control horizon imposed by the \gls{cb}'s timescale. During a 24-hour control experiment, we run the convex \gls{mpc} implemented in Section~\ref{Flat-rate_tariff}. For 30\% of the iterations, we run the solver for 5, 10, and 30 minutes. We find that in 92.4\% of the instances, the same solution was found with a computation time of 5 minutes as in 10 minutes, and in 78.3\% when compared to a computation time of 30 minutes.
Figure~\ref{fig:optimiality_test} illustrates the absolute gap between the objective \textcolor{blue}{function} values, $\vert \Delta f \vert$. To contextualize these results, large gaps can be attributed to variations in peak power between solutions, as peak power demand is strongly penalized by a rate of \$14.58/kW.  However, it's important to note that the objective \textcolor{blue}{function} values do not reflect real operational costs. This is because we optimize peak power over the \gls{mpc} horizon rather than considering the observed monthly peak.





\section{Conclusion}\label{conclusion}
In this work, we propose an energy management method for \gls{rtu}-\gls{hvac} of small \glspl{cb}. Our convex \gls{mpc} approach levrages a discrete and data-driven model of the thermal dynamics, making the implementation in \gls{rtu}-based \glspl{cb} straightforward. We develop a bidding strategy for the real-time market and apply our approach to a \gls{tou} and a \gls{cpr} program. The performance of the proposed controller is evaluated through a numerical building simulation, and has demonstrated a good compromise between thermal comfort, energy reduction, and cost. In future work, we will extend our approach to an aggregation of small CBs \textcolor{red}{or to a large CB through distributed optimization via the \gls{admm}~\cite{MAL-016}}. We also plan to propose a resolution method that produces provably optimal solutions. 


\bibliographystyle{IEEEtran}
\bibliography{reference.bib}


 




\vfill

\end{document}